\documentclass[a4paper,11pt]{article}
\usepackage{geometry}
\usepackage{a4wide}
\usepackage{graphicx}
\usepackage{epsf}
\usepackage{amsmath}
\usepackage{amssymb}
\usepackage[normalem]{ulem}

\usepackage{cite}
\usepackage{multirow}
\usepackage{subcaption}
\usepackage{appendix}
\usepackage{tikz}
\usetikzlibrary{calc}
\usepackage{subcaption}
\usepackage{braket}
\usepackage{amsfonts,amsthm,euscript,braket,xcolor}
\usepackage{breqn}
\usepackage[colorlinks=true, linkcolor=blue, citecolor=blue, urlcolor=blue]{hyperref}  

\newcommand{\be}{\begin{equation}}
	\newcommand{\ee}{\end{equation}}

\newcommand{\Rmnum}[1]{\expandafter\@slowromancap\romannumeral #1@}
\newcommand{\bea}{\begin{eqnarray}}
	\newcommand{\eea}{\end{eqnarray}}

\numberwithin{equation}{section}

\usepackage{ulem}

\begin{document}
	
	\title{\bf A note on the holographic time-like entanglement entropy in Lifshitz theory}
	\author{ \textbf{\textbf{Siddhi Swarupa Jena}\thanks{siddhiswarupa\_jena@nitrkl.ac.in}, \textbf{Subhash Mahapatra}\thanks{mahapatrasub@nitrkl.ac.in} } 
			\\\\\textit{{\small Department of Physics and Astronomy, National Institute of Technology Rourkela, Rourkela - 769008, India}}
			}
	\date{}
		\maketitle
	
	\begin{abstract}
		We explore the holographic time-like entanglement entropy (TEE) in the boundary theory of three-dimensional Lifshitz spacetime. There have been various holographic proposals for TEE in recent years and we test those proposals in the Lifshitz background.  We obtain the analytic result for TEE in each proposal, compare the results, and analyze how the anisotropic scaling affects the TEE. We find that different holographic proposals give the same result for TEE in the Lifshitz background. Our analysis further suggests that the TEE of the Lifshitz system contains real and imaginary parts, both of which depend on the anisotropic parameter.   
  \end{abstract}
	
	\section{Introduction}
	\label{sec1}
The holographic principle has profoundly transformed our understanding of the relationship between quantum field theories and gravity. Its most explicit realization, the AdS/CFT correspondence\cite{Maldacena:1997re,Witten:1998qj,Gubser:1998bc}, posits a duality between a gravitational theory in a ($d+1$)-dimensional anti-de Sitter (AdS) space and a conformal field theory (CFT) living in its $d$-dimensional boundary. This duality has proven to be an indispensable tool for studying strongly coupled quantum theories, which are usually difficult to study through traditional perturbative techniques. This duality depicts that the geometry of the bulk AdS space should be encoded in the quantum properties of CFT. This relationship is crucial for understanding how spacetime geometry can arise from field-theoretic principles \cite{VanRaamsdonk:2010pw,Balasubramanian:2013lsa}. A key quantitative tool for exploring this emergent geometry is the holographic entanglement entropy (HEE). In this context, the entanglement entropy of a spatial region in CFT is computed as the area of a corresponding extremal surface embedded in the AdS spacetime, in accordance with the Ryu-Takayanagi prescription \cite{Ryu:2006bv,Hubeny:2007xt,Ryu:2006ef}
\begin{equation}
    S_A = \frac{Area(\Gamma_A)}{4 G_{d+1}},
\end{equation}
where $\Gamma_A$ is the bulk extremal surface whose boundary is homologous to the boundary of the subsystem $A$. This prescription provides a concrete link between quantum entanglement in the field theory and geometric properties in the dual gravitational theory. In the last two decades, the space-like entanglement measure has provided valuable insights into the structure of spacetime, quantum field theory, black hole physics, etc. See \cite{Ryu:2006ef,Rangamani:2016dms}, for reviews of the holographic entanglement entropy and its application in strongly coupled field theories.

In recent years, there has been growing interest in studying time-like entanglement entropy (TEE) in holography and condensed matter systems \cite{Liu:2022ugc,Doi:2022iyj,Doi:2023zaf}.   The TEE considers entanglement between regions separated in time, unlike space-like entanglement, which examines spatial correlations. The TEE contains imaginary contributions and, therefore, is a pseudo-entropy rather than a von Neumann entropy. In holography, to connect the boundary of a time-like subsystem one usually combines different extremal surfaces, such as space-like and time-like, to compute the corresponding TEE. This is because, usually, there is no definite space-like or time-like extremal surface that is homologous to the time-like subsystem at the AdS boundary. In \cite{Doi:2022iyj,Doi:2023zaf}, TEE in two-dimensional CFT was obtained by analytical continuation of a space-like interval to a time-like interval. It was shown that the same TEE expression could be obtained holographically by joint space-like and time-like extremal surfaces in bulk, i.e.,  the space-like and time-like surfaces respectively give real and imaginary contributions to TEE. For further related works on TEE, one may refer to \cite{Reddy:2022zgu, Narayan:2022afv,  Giudice:2021smd, Olson:2011bq, He:2024emd, Anegawa:2024kdj, Basu:2024bal, Guo:2024lrr, Das:2023yyl, Grieninger:2023knz, Narayan:2023zen, He:2023ubi, Chu:2023zah, Jiang:2023loq, Narayan:2023ebn, Chen:2023gnh, Caputa:2024gve}.

It is desirable and expected that for a holographic definition of TEE to be well-defined, it should exist independent of the analytical continuation. In particular, it is not always possible to perform analytic continuation of the space-like subsystem in general dual field theories, as this requires a complete analytic expression of the space-like extremal surfaces, which may not be readily available in more complicated gravity theories. For these reasons, there have been various proposals for the holographic TEE in recent years \cite{Li:2022tsv,Afrasiar:2024lsi,Heller:2024whi}. In \cite{Li:2022tsv},  TEE was suggested to be given by the complex-valued weak extremal surface (CWES), a complex-valued generalization of the Ryu-Takayanagi extremal surfaces. This was based on the idea that there can in fact be multiple combined space-like and time-like surfaces that are homologous to the boundary subsystem and one therefore should implement CWES criteria to obtain TEE (see section~\ref{sec3} for more details). The proposal of \cite{Afrasiar:2024lsi} is based on the idea of the smooth merging of space-like and time-like surfaces by having a well-defined first derivative of the space-like and time-like surfaces at the merging point. Similarly, the proposal \cite{Heller:2024whi} introduced a novel idea of complexifying the spacetime coordinates with TEE given by the Ryu-Takayanagi extremal surface in this complexified bulk geometry.   

The AdS$_3$/CFT$_2$ duality is generally used as a bona fide tabletop laboratory to test the correctness of any new holographic proposal. This is mainly because in pure AdS$_3$ extremal surfaces just become geodesics in the bulk. Therefore analytic results are relatively easier to obtain in AdS$_3$/CFT$_2$ settings, hence providing a feasible scenario to get a comprehensive understanding of the proposal. In this context, the above-mentioned seemingly different holographic proposals correctly reproduce the TEE result of CFT$_2$, giving credit to their prescription. Once the proposal is verified one can use it to investigate more complicated systems, such as non-conformal theories, thermal systems having black holes in the bulk, anisotropic systems, etc.

While the AdS$_3$/CFT$_2$ case has been well-explored, with known analytical results for TEE, the study of TEE in more complex gravitational backgrounds remains the frontier in holographic research. To test and further scrutinize various holographic TEE proposals it is important to investigate them in different gravitational backgrounds and see if they again lead to the same result and conclusion. It would also be desirable if the TEE computation is possible in an analytic manner. This may not only allow us to see the strengths and weaknesses of each proposal explicitly but may also help us find similarities and universality between them. In this work, we use one such gravitational system, namely the three-dimensional Lifshitz background, and obtain TEE of the dual boundary theory analytically using the above-mentioned holographic approaches.

Lifshitz field theories are a class of quantum field theories that exhibit anisotropic scaling between space and time. Holographically, Lifshitz field theories are dual to Lifshitz spacetimes, which break Lorentz symmetry by having different scaling between space and time coordinates \cite{Kachru:2008yh,Taylor:2008tg}. The metric for a Lifshitz spacetime in three dimensions is typically written as,
\begin{equation}
     ds^2= L^2 \left[ -\frac{dt^2}{z^{2\alpha}}+\frac{dz^2}{z^2}+\frac{dx^2}{z^2}\right]\,,
    \label{lifshitzmetric}
\end{equation}
where $z$ is the holographic radial coordinate in the bulk, $x$ is the spatial coordinate, and $\alpha$ is the critical exponent characterizing the anisotropic scaling. The Lifshitz metric is invariant under the scaling symmetry
\begin{equation}
t\rightarrow \lambda^\alpha t,~~z\rightarrow \lambda z,~~x\rightarrow \lambda x\,,
\end{equation}
which is consistent with the Lifshitz symmetry of the dual field theory \cite{Kachru:2008yh}. The metric of the form (\ref{lifshitzmetric}) is generally a solution to the Einstein equation
with additional matter fields, such as $p$-form gauge fields \cite{Kachru:2008yh}. The reality condition of the corresponding fluxes requires $\alpha \geq 1$. In the last few years, Lifshitz spacetimes have been thoroughly used to discuss strong coupling physics holographically in non-relativistic settings, see \cite{Taylor:2015glc} for a review. Note that at constant time, the metric (\ref{lifshitzmetric}) takes the same form as that for Pure AdS$_3$. Accordingly, the holographic entanglement entropy of the dual field theory for a space-like interval would be the same for these two backgrounds.\footnote{For a discussion related to the space-like entanglement entropy in two-dimensional Lifshitz theory based on holography inspired continuum multi-scale entanglement renormalization ansatz (cMERA), see \cite{Nozaki:2012zj,He:2017wla,Gentle:2017ywk}. See \cite{Vasli:2024mrf, Wang:2021qey,Mozaffar:2021nex,MohammadiMozaffar:2017nri}, for works on the space-like entanglement structure in Lifshitz theories.}  However, for a time-like interval, this is not the case and one is expected to get $\alpha$ dependent anisotropic TEE, thereby providing another important reason to investigate TEE in Lifshitz theories holographically.      

In this paper, we investigate TEE in two-dimensional Lifshitz theory holographically using the formalism suggested in \cite{Doi:2023zaf,Li:2022tsv,Afrasiar:2024lsi,Heller:2024whi}. We will consider $\alpha \geq 1$ as in \cite{Kachru:2008yh}. We obtain the analytic result for TEE in each formalism, compare the results, and analyze how the anisotropic scaling affects the TEE. We find that all these holographic formalisms give the same results for TEE in the Lifshitz background as well. Both the real and imaginary parts of TEE now depend on $\alpha$ and reduce to the CFT$_2$ expression in the limit $\alpha$ goes to one. This suggests Lifshitz field theories that are characterized by a different anisotropic parameter $\alpha$ exhibit different TEE structures. Similar to the CFT$_2$ case, the imaginary part of TEE does not depend on subsystem length but relies on $\alpha$.

The structure of the paper is outlined as follows. In Section~\ref{sec2} we study the holographic TEE via geodesics in the three-dimensional Lifshitz geometry. We compute TEE by the complex-valued extremal surface (CWES) method in Section~\ref{sec3}. Section~\ref{sec4} is dedicated to calculating TEE via smooth merging of time-like and space-like geodesics. Finally, in Section~\ref{sec5}, we discuss our final method to calculate TEE holographically by coordinate complexification. Lastly, Section~\ref{sec6} encompasses our primary findings
and provide an outlook for possible future research directions.

\section{HTEE via Geodesics: Method~1}
\label{sec2}
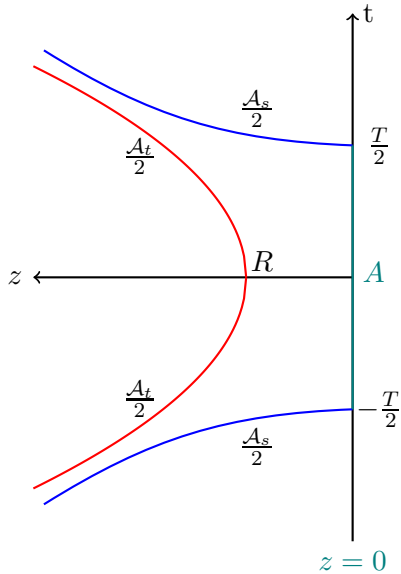
\begin{figure}[htb!]
  \centering 
\begin{tikzpicture}[scale=0.7,rotate=180]
    \draw[thick,->](0,0) -- (6,0) node[left] {$z$};
    \draw[thick,<-](0,-5) -- (0,5) ;

\draw[blue,thick] (0,-2.5) to[bend left=15] (5.8,-4.3);
\draw[blue,thick] (0,2.5) to[bend right=15] (5.8,4.3);
    \draw[thick,red] plot[domain=1:5, samples=100]({\x+1},{2*sqrt(\x-1)});
    \draw[thick,red] plot[domain=1:5, samples=100]({\x+1},{-2*sqrt(\x-1)});

    \node at (-0.5,2.5) {$-\frac{T}{2}$};
    \node at (-0.5,-2.5) {$\frac{T}{2}$};
    \node at (-0.4,-0.1) {\textcolor{teal}{$A$}};
    \node at (1.7,-0.3) {$R$};
    \node at (4,2.3) {$\frac{\mathcal{A}_t}{2}$};
    \node at (4,-2.3) {$\frac{\mathcal{A}_t}{2}$};
    \node at (1.8,3.2) {$\frac{\mathcal{A}_s}{2}$};
    \node at (1.8,-3.2) {$\frac{\mathcal{A}_s}{2}$};
    \node at (-0.3,-5) {t};
    \node[below] at (0,5) {\textcolor{teal}{$z=0$}};
    \draw[teal,thick] (0,2.5) -- (0,-2.5);
\end{tikzpicture}
\caption{ Geodesics related to the computation of holographic TEE in the Poincar\'{e} patch. The blue curves correspond to space-like geodesics and give real contributions to TEE, whereas the red curves correspond to time-like geodesics and give imaginary contributions.}
\label{geodesicsmethod1}
\end{figure}

To set the stage for later sections, in this section, we first reproduce the results of \cite{Doi:2023zaf,Basak:2023otu} concerning the holographic time-like entanglement entropy in the two-dimensional field Lifshitz theory. In three dimensions, the time-like (or the spatial) entanglement entropy is related to the length of the geodesics connecting the endpoints of the subsystem.  For a time-like strip subsystem $A$ on the boundary defined by
\begin{equation}
    \tau = \{(t,x):t \in \left[-\frac{T}{2},\frac{T}{2}\right],x=0\}\,,
    \label{strip}
\end{equation}
the holographic TEE consists of two space-like geodesics that propagate from the endpoints of $A$ to the future and past infinities plus a time-like geodesic that connects the endpoints of two space-like geodesics \cite{Doi:2022iyj,Doi:2023zaf}. A pictorial representation of these geodesics is given in Fig.~\ref{geodesicsmethod1}, where the blue curves correspond to the space-like geodesics and a red curve corresponds to the time-like geodesics. Interestingly, in the Poincar\'{e} patch of the Lifshitz spacetime, one can find analytic expressions of these geodesics in the $t-z$ plane. These are given by  
\begin{eqnarray}
t & = & \frac{1}{\alpha}\sqrt{z^{2\alpha}+\frac{\alpha^2 T^2}{4}},~~~~~~~ \text{space-like}
\label{spacelikegeo}\\
z^{2\alpha} & = & \alpha^2 (t-t_0)^2 + R^{2\alpha},~~~~~ \text{time-like}
\label{timelikegeo}
\end{eqnarray}
where $t_0$ and $R$ are two constants, which can be determined from the positions of the endpoints of the time-like geodesics. As we will see shortly, though positions of the endpoints of the time-like geodesics depend on constants $t_0$ and $R$, its length does not. 

Using Eqs.~(\ref{spacelikegeo}) and (\ref{timelikegeo}), one can compute the lengths of the space-like and time-like geodesics. For the space-like geodesics, since they extend from the asymptotic boundary to the future and past null infinities, their total length is given by
\begin{eqnarray}
\mathcal{A}_s=2 L \int_{\epsilon}^{\infty} dz~\sqrt{\frac{1}{z^2}-\frac{t'(z)^2}{z^{2\alpha}} }\,,
\end{eqnarray}
where $\epsilon$ is a UV cutoff and the prime $'$ denotes the derivative with respect to $z$. Substituting $t'(z)$ from Eq.~(\ref{spacelikegeo}), we get
\begin{eqnarray}
\mathcal{A}_s &=& 2 L \int_{\epsilon}^{\infty} dz~ \frac{\alpha T}{2} \frac{1}{z \sqrt{z^{2\alpha} + \frac{\alpha^2 T^2}{4}}} =\frac{2 L}{\alpha} \sinh^{-1}\left(\frac{\alpha T}{2 \epsilon^\alpha}\right) \,.
\end{eqnarray}
In the limit $\epsilon \rightarrow 0$, we get
\begin{eqnarray}
\mathcal{A}_s &=& \frac{2 L}{\alpha}\log{\frac{\alpha T}{\epsilon^\alpha}} + \mathcal{O}(\epsilon^\alpha) \,,
\label{spacelikelengthM1}
\end{eqnarray}
where we have used the relation  $\sinh^{-1}x=\log(x+\sqrt{1+x^2})$. In the limit $\epsilon \rightarrow 0$, the dominant contribution comes from the first term $ \frac{2 L}{\alpha}\log{\frac{\alpha T}{\epsilon^\alpha}}$. For simplicity, we will neglect higher-order terms in $\epsilon$ in the subsequent analysis.

Similarly, the length of time-like geodesics is computed as\footnote{We will see in the next section that the time-like geodesic goes from the past null infinity ($t=-\infty$) to the future null-infinity ($t=+\infty$).} 
\begin{eqnarray}
\mathcal{A}_t= 2 i L R^\alpha \int_{R}^{\infty} dz~ \frac{1}{z \sqrt{z^{2\alpha} - R^{2\alpha}}} = i \frac{\pi L}{ \alpha} \,.
\label{timelikelengthM1}
\end{eqnarray}
Combining (\ref{spacelikelengthM1}) and (\ref{timelikelengthM1}), the TEE of the Lifshitz field theory in vacuum is given by 
\begin{eqnarray}
 S_{A}^{T} & = & \frac{1}{4 G_3}\left( \mathcal{A}_s + \mathcal{A}_t  \right)\,, \nonumber\\
 & = &  \frac{L}{4 G_3}\left( \frac{2}{\alpha}\log{\frac{\alpha T}{\epsilon^\alpha}} +  i \frac{\pi}{ \alpha}  \right)\,.
\label{TEEM1}
\end{eqnarray}
Note that the above TEE expression reduces to the usual TEE expression of the two-dimensional CFT in the limit $\alpha$ goes to one. This is expected since for $\alpha=1$, the Lifshitz geometry reduces to pure AdS$_3$. Moreover, both real and imaginary parts of the TEE now depend on the anisotropic parameter $\alpha$, particularly the imaginary part which is inversely proportional to $\alpha$. Therefore, Lifshitz field theories with different values of the anisotropic parameter are expected to have different imaginary TEE parts. The same is true for the real part, however with slightly more complicated dependence. Like in the case of CFT$_2$, the imaginary part of TEE is constant and does not depend on the subsystem size. Unlike the real part, the imaginary part does not contain any UV divergences and is UV finite in
nature. Note also that the above TEE expression differs from the one in \cite{Basak:2023otu}. In particular, the argument of the logarithmic term in the real part of TEE is $\alpha T$ instead of just $T$. Later, we will see that the same $\alpha T$ dependence is observed in the real part with different methods.

\section{HTEE via CWES: Method~2}
\label{sec3}
\begin{figure}
    \centering
   \begin{tikzpicture}
\path (7,7) coordinate (A)
    (7,0) coordinate (B)
    (3,3.5) coordinate (C)
    ($(A)!0.25!(B)$) coordinate (A1) node[right] {$A_1$}
    ($(A)!0.3!(C)$) coordinate (A2) node[above left] {$A_2$}
    ($(A)!0.6!(C)$) coordinate (A3) node[above left] {$A_3$}
    ($(B)!0.25!(A)$) coordinate (B1) node[right] {$B_1$}
    ($(B)!0.3!(C)$) coordinate (B2)  node[below left] {$B_2$}
    ($(B)!0.6!(C)$) coordinate (B3)  node[below left] {$B_3$};
\draw[blue,thick] (A1) to[bend left=20] (A2)
    (A1) to[bend left=20] (A3)
    (B1) to[bend right=20] (B2)
    (B1) to[bend right=20] (B3);
\draw[red,thick] (A2) to[bend right=30] (B2)
    (A3) to[bend left=10] (B3);
\draw[line width=1.5pt] (A) -- (B) -- (C) -- cycle;
\end{tikzpicture}
    \caption{Penrose diagram on  Poincar\'{e} patch of the Lifshitz spacetime and two possible geodesics configurations connecting the time-like interval $A_1 B_1$}
    \label{CWESmethod}
\end{figure}
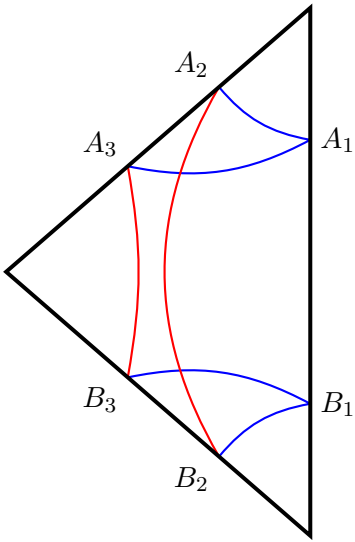

In this section, we compute TEE in the Lifshitz field theory using the formalism suggested in \cite{Li:2022tsv}. The essential idea here is that multiple bulk surfaces can be homologous to the boundary subsystem if one allows for the possibilities of combined piecewise space-like and time-like extremal surfaces.  Two of such surfaces are shown in Fig.~\ref{CWESmethod}. The areas of these multiple surfaces can be different, leading to different values of TEE from holography. To illustrate this point, note that the geodesic equation for the Lifshitz geometry takes the form
\begin{eqnarray}
& & z \ddot z + (2\alpha-1)\dot z^2 -\alpha z^{2-2\alpha}=0 \,,
\end{eqnarray}
where the dot indicates the derivative with respect to $t$. Its general space-like solution is 
\begin{eqnarray}
& & t= \frac{1}{\alpha}\sqrt{z^{2\alpha}+C_{1}^{2}}+C_2,~~~t=-\frac{1}{\alpha}\sqrt{z^{2\alpha}+C_{1}^{2}}-C_2 \,.
\end{eqnarray}
Therefore, the general solution has two branches, each having two undetermined constants $C_1$ and $C_2$. If we choose the same constants $\{C_1=\alpha T/2, ~C_2=0\}$ for both branches, the above geodesic solutions reduce to Eq.~(\ref{spacelikegeo}), i.e.,  Eq.~(\ref{spacelikegeo}) corresponds to a particular space-like solution to the geodesic equation. For these particular values of constants, one gets the TEE as in Eq.~(\ref{TEEM1}). However, as far as the homologous condition of the bulk geodesic and boundary subsystem is concerned, the above geodesic solutions are not unique. There are other possibilities for geodesics that respect the homologous requirement with the subsystem boundary. In general, the two branches of the space-like geodesic could be 
\begin{eqnarray}
& & t= \frac{1}{\alpha}\sqrt{z^{2\alpha}+C_{+}^{2}}-\frac{C_{+}}{\alpha}+\frac{T}{2},~~~t=-\frac{1}{\alpha}\sqrt{z^{2\alpha}+C_{-}^{2}}+\frac{C_{-}}{\alpha}-\frac{T}{2}\,,
\label{geodesicgeneral}
\end{eqnarray}
where $C_{\pm}$ are two constants which, without losing any generality, we take to be non-negative. Note that at the asymptotic boundary $z=0$, these two branches match the endpoints of $\partial A$. Interestingly, the length of the space-like geodesic, and therefore TEE, depends on constants $C_{\pm}$. In particular, the total length of the two space-like geodesics is now given by
\begin{eqnarray}
& & \mathcal{A}_s =  \frac{1}{4 G_3 }\frac{L}{\alpha}\log{\frac{4C_{+}C_{-}}{\epsilon^{2\alpha}}}\,.
\label{spacelikeM2general}
\end{eqnarray}
If we agree that the two branches of the space-like geodesic go to future infinity ($t\rightarrow +\infty, z\rightarrow +\infty$)  and past infinity ($t\rightarrow -\infty, z\rightarrow +\infty$), the equation for the time-like geodesic connecting these two infinities would be the same as in (\ref{timelikegeo}). Accordingly, the length of the time-like geodesic would again be imaginary and take the value $i\pi L/\alpha$. Therefore, TEE in general will have the expression  
\begin{eqnarray}
& & S_{A}^{T} =  \frac{1}{4 G_3 }\frac{L}{\alpha}\left(\log{\frac{4C_{+}C_{-}}{\epsilon^{2\alpha}}} + i \pi \right)\,.
\label{TEEM2general}
\end{eqnarray}
We see that different $C_{\pm}$ gives different TEE for the two-dimensional Lifshitz field theory.

Notice that different pair of $C_{\pm}$ corresponds to different space-like+time-like geodesics that are homologous to $\partial A$ at the asymptotic boundary. Accordingly, it raises the question of how to correctly choose a particular value of $C_{\pm}$. In the context of AdS$_3$/CFT$_2$, this can be done by demanding the holographic expression to match with the dual CFT expression which is obtained by analytic continuation of the space-like interval to the time-like interval. However, such analytic continuation and TEE results are not available for field theories in general, thereby causing fixing of $C_{\pm}$ a complicated affair.    

To address these issues, a notion of complex-valued weak extremal surface (CWES), which is a complex-valued generalization of the Ryu-Takayanagi extremal surfaces, was suggested for TEE in \cite{Li:2022tsv}. To understand CWES, let $\mathcal{X}_{\mathcal{B}}$ is a set of all dimension-two piecewise smooth surface $\Gamma=\bigcup_{i}^{N}\Gamma_i$, where $\Gamma_i$ are individual smooth surfaces. The surfaces $\Gamma_i$ can join together at the joint point marked as $E_{ij}=\Gamma_i\bigcup \Gamma_j$. One calls $\Gamma\in \mathcal{X}_{\mathcal{B}}$ to be a CWES if (i) it's every individual piece $\Gamma_i$ is an extremal surface, (ii) in case there are multiple extremal surfaces $\Gamma_i$ enclosed by $\partial \Gamma_i$, then $\Gamma_i$ is the one which has the minimal area, and (iii) the area of $\Gamma$ will be functional of joints and should be extremal with respect to them, i.e.,  
\begin{eqnarray}
& & \frac{\delta \mathcal{A}(\Gamma)}{\delta E_{ij}} = 0\,.
\label{jointcondition}
\end{eqnarray}
With CWES, the generalized holographic entanglement entropy proposal is
\begin{eqnarray}
& & S_{A}^{T} =  \text{Min} \Bigg\{ \mathop{Ext}_{\partial \Gamma= \partial A} \frac{\mathcal{A}(\Gamma)}{4G_D} \Bigg\}\,,
\label{TEEproposalCWES}
\end{eqnarray}
where $Ext$ stands for the complex-valued weak extremal surfaces, $\mathcal{A}$ is the complex-valued area of the piecewise surface $\Gamma=\bigcup_{i}^{N}\Gamma_i$. In principle, there can be many CWES, and the formalism is to take that CWES, which has the lowest imaginary area. In case two CWES have the same imaginary area, then take the CWES that has the lowest real area. 

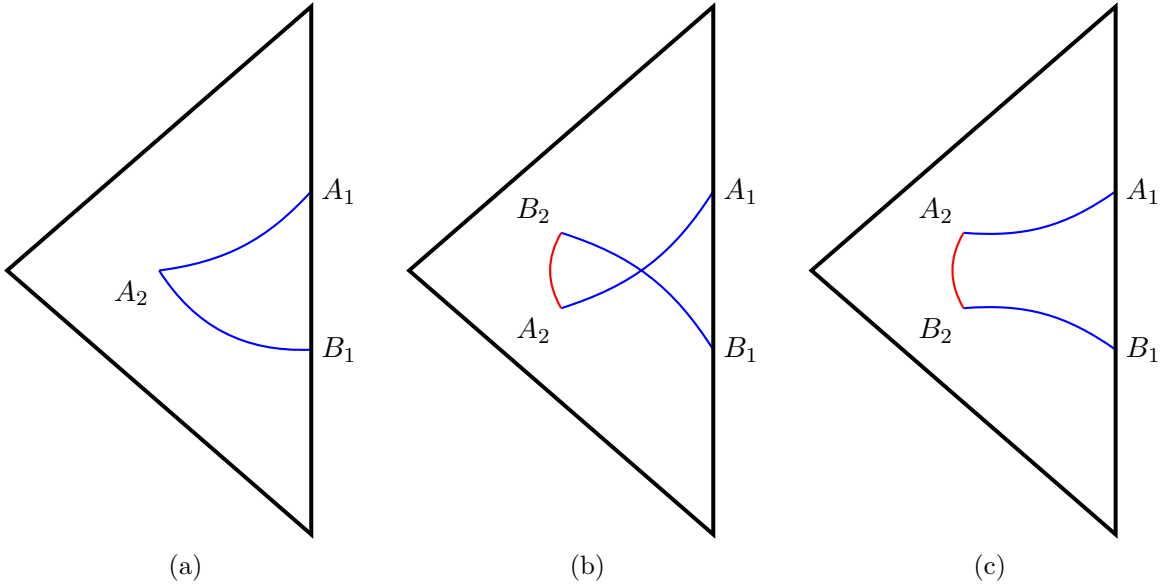
\begin{figure}[h]
  \centering
     \begin{minipage}{0.32\textwidth}
        \centering
        \begin{tikzpicture}
            \path (7,7) coordinate (A)
            (7,0) coordinate (B)
            (3,3.5) coordinate (C)
            ($(A)!0.35!(B)$) coordinate (A1) node[right] {$A_1$}
            ($(B)!0.35!(A)$) coordinate (B1) node[right] {$B_1$}
            (5,3.5) coordinate (A2)  node[below left] {$A_2$};
            \draw[blue,thick] (A1) to[bend left=20] (A2);
            \draw[blue,thick] (B1) to[bend left=30] (A2);
            \draw[line width=1.5pt] (A) -- (B) -- (C) -- cycle;
        \end{tikzpicture}
        \subcaption{}
        \label{CWES1}
    \end{minipage}
    \hfill
    \begin{minipage}{0.32\textwidth}
        \centering
        \begin{tikzpicture}
            \path (7,7) coordinate (A)
            (7,0) coordinate (B)
            (3,3.5) coordinate (C)
            ($(A)!0.35!(B)$) coordinate (A1) node[right] {$A_1$}
            (5,4) coordinate (B2) node[above left] {$B_2$}
            ($(B)!0.35!(A)$) coordinate (B1) node[right] {$B_1$}
            (5,3) coordinate (A2)  node[below left] {$A_2$};
            \draw[blue,thick] (A1) to[bend left=20] (A2)
            (B1) to[bend right=20] (B2);
            \draw[red,thick] (A2) to[bend left=30] (B2);
            \draw[line width=1.5pt] (A) -- (B) -- (C) -- cycle;
        \end{tikzpicture}
        \subcaption{}
        \label{CWES2}
    \end{minipage}
    \hfill
    \begin{minipage}{0.32\textwidth}
        \centering
        \begin{tikzpicture}
            \path (7,7) coordinate (A)
            (7,0) coordinate (B)
            (3,3.5) coordinate (C)
            ($(A)!0.35!(B)$) coordinate (A1) node[right] {$A_1$}
            (5,4) coordinate (A2) node[above left] {$A_2$}
            ($(B)!0.35!(A)$) coordinate (B1) node[right] {$B_1$}
            (5,3) coordinate (B2)  node[below left] {$B_2$};
            \draw[blue,thick] (A1) to[bend left=20] (A2)
            (B1) to[bend right=20] (B2);
            \draw[red,thick] (A2) to[bend right=30] (B2);
            \draw[line width=1.5pt] (A) -- (B) -- (C) -- cycle;
        \end{tikzpicture}
        \subcaption{}
        \label{CWES3}
    \end{minipage}
    \caption{Three different possible geodesic representations to connect the endpoint of time-like interval $A_1 B_1$.}
    \label{diffgeodesicsCWES}
\end{figure}

Let us now apply the above proposal to the Lifshitz background in three dimensions in the Poincar\'{e} patch. The computation of TEE here is completely analogous to the AdS$_3$ case \cite{Li:2022tsv}. Just like in AdS$_3$, there is no smooth geodesic to connect the two endpoints of the time-like subsystem $A$ at the boundary $z=0$. Accordingly, CWES will be made of a few piecewise smooth extremal geodesics. The translation symmetry along the $x$ direction ensures that there are only three possibilities to combine space-like and time geodesics which are homologous to the boundary of the subsystem $A$.\footnote{One can also use four null geodesics to connect the endpoints of $A$. However, such a configuration is not a CWES.}  These possibilities are illustrated in Fig.~\ref{diffgeodesicsCWES}, where blue and red curves correspond to space-like and time-like geodesics, respectively.  Note that in general the joints can exist anywhere in the bulk and not necessarily at the future and past null infinities. Shortly, we will see that the requirement of CWES forces these joints to be only at future and past null infinities. 

Let us now first analyze the geodesic structure (\ref{CWES1}). The space-like geodesics $A_1A_2$ and $B_1A_2$ are described by 
\begin{eqnarray}
& & A_1A_2:~~ t= \frac{1}{\alpha}\sqrt{z^{2\alpha}+C_{+}^{2}}-\frac{C_{+}}{\alpha}+\frac{T}{2},
\label{spacelike1CWEScase1}
\end{eqnarray}
\begin{eqnarray}
& & B_1A_2 :~~~t=-\frac{1}{\alpha}\sqrt{z^{2\alpha}+C_{-}^{2}}+\frac{C_{-}}{\alpha}-\frac{T}{2}\,.
\label{spacelike2CWEScase1}
\label{geodesicgeneral}
\end{eqnarray}
Taking the bulk position of $A_2=(t_2,z_2)$, the total length of the two space-like geodesics $A_1A_2$ and $B_1A_2$ is given by
\begin{eqnarray}
& & \mathcal{A}_s =  L \left[ \int_{\epsilon}^{z_2} dz~\frac{C_{+}}{z\sqrt{z^{2\alpha}+C_{+}^{2}}} + \int_{\epsilon}^{z_2} dz~ \frac{C_{-}}{z\sqrt{z^{2\alpha}+C_{-}^{2}}}  \right]\,,
\label{CWEStotallengthcase1}
\end{eqnarray}
where $z_2$ is a function of $C_{\pm}$. The joint condition $\partial \mathcal{A}/\partial C_{\pm}=0$ of CWES leads to
\begin{eqnarray}
& & \left[ \frac{C_{+}}{z_2\sqrt{z_{2}^{2\alpha}+C_{+}^{2}}} + \frac{C_{-}}{z_2\sqrt{z_{2}^{2\alpha}+C_{-}^{2}}}\right]\frac{\partial z_2}{\partial C_{\pm}}=0\,.
\label{combinespacegeoM2Cwes1}
\end{eqnarray}
However, at $A_2$ we also have [from Eq.~(\ref{geodesicgeneral})]
\begin{eqnarray} 
& & \sqrt{z_{2}^{2\alpha}+C_{+}^{2}} + \sqrt{z_{2}^{2\alpha}+C_{-}^{2}}=C_{+}+C_{-}-\alpha T\,,
\end{eqnarray}
which leads to
\begin{eqnarray}
& & \left[ \frac{C_{+}}{z_2\sqrt{z_{2}^{2\alpha}+C_{+}^{2}}} + \frac{C_{-}}{z_2\sqrt{z_{2}^{2\alpha}+C_{-}^{2}}}\right]\frac{\partial z_2}{\partial C_{\pm}}=\frac{C_{\pm}}{\alpha z_2} \left[ 1- \frac{C_{\pm}}{\sqrt{z_{2}^{2\alpha}+C_{\pm}^{2}}} \right] > 0\,
\label{combinespacegeoM2Cwes2},
\end{eqnarray}
which implies that Eqs.~(\ref{combinespacegeoM2Cwes1}) and (\ref{combinespacegeoM2Cwes2}) can not be simultaneously satisfied. Accordingly, the total length (\ref{CWEStotallengthcase1}) is not extremal, and the geodesic structure (\ref{CWES1}) is ruled out for TEE. This is also expected, considering that it is not possible to connect two time-like boundary points by space-like bulk geodesics alone \cite{He:2024emd}. 

Now let us turn our attention to the geodesic structure (\ref{CWES2}). In this case, the space-like geodesics $A_1A_2$ and $B_1 B_2$ are described by 
\begin{eqnarray}
& & A_1A_2 :~~~t= -\frac{1}{\alpha}\sqrt{z^{2\alpha}+C_{+}^{2}}+\frac{C_{+}}{\alpha}+\frac{T}{2},
\label{spacelike2case2}\\
&  & B_1 B_2 :~~~t= \frac{1}{\alpha}\sqrt{z^{2\alpha}+C_{-}^{2}}-\frac{C_{-}}{\alpha}-\frac{T}{2}\,,
\label{spacelike3case2}
\end{eqnarray}
whereas, the time-like geodesic $A_2B_2$ is again described by 
\begin{eqnarray}
& & z^{2\alpha}=\alpha^2 (t-t_0)^2 + R^{2\alpha}\,.
\label{timelikeCWEScase2}
\end{eqnarray}
Taking the bulk coordinates of points $A_2$ and $B_2$ as
\begin{eqnarray}
& & A_2=(t_2,z_2),~~~B_2=(t_3,z_3),~~~\text{with}~~t_3>t_2\,,
\end{eqnarray}
the length of the time-like geodesic $A_2B_2$ is
\begin{eqnarray}
& & \mathcal{A}_t = L \int_{t_2}^{t_3} dt~\sqrt{-\frac{1}{z^{2\alpha}}+\frac{\dot z^2}{z^2}} = i L \int_{t_2}^{t_3} dt \frac{R^\alpha}{\alpha^2 (t-t_0)^2 + R^{2\alpha}},   \nonumber \\
& & = \frac{i L}{\alpha} \left[\tan^{-1}\left(\frac{\alpha (t_3-t_0)}{R^\alpha} \right) - \tan^{-1}\left(\frac{\alpha (t_2-t_0)}{R^\alpha} \right)  \right]
\label{lengthtimelikeCWES2}
\end{eqnarray}
The joint condition (\ref{jointcondition}) requires 
\begin{eqnarray}
& & \frac{\partial \mathcal{A}_t}{\partial t_2}=0,~~~\frac{\partial \mathcal{A}_t}{\partial t_3}=0\,.
\label{jointequationCWEScase2}
\end{eqnarray}
This gives us $t_2=-\infty$ and $t_3=\infty$. Now, expanding Eqs.~(\ref{spacelike3case2}) and (\ref{timelikeCWEScase2}) near the future null infinity, we get
\begin{eqnarray}
& & z^\alpha = \alpha(t-t_0)+\mathcal{O}(1/t),~~~t=\frac{z^\alpha}{\alpha}-\frac{C_{-}}{\alpha}-\frac{T}{2}+\mathcal{O}(1/z^\alpha)\,.
\end{eqnarray}
These two equations represent the same point $A_2$ when $z\rightarrow \infty$. Accordingly, we must have $t_0=-C_{-}/\alpha -T/2$. Doing the analogous expansion for the point $B_2$, we get $t_0=C_{+}/\alpha + T/2$. Therefore, we conclude that
\begin{eqnarray}
& & C_{+} + C_{-}=-\alpha T\,.
\end{eqnarray}
This violates the assumption that the constants $C_{\pm}$ are nonnegative. Therefore, the joint equation (\ref{jointequationCWEScase2}) has no solution and the structure (\ref{CWES2}) is not a CWES. Accordingly, the structure (\ref{CWES2}) is not appropriate for the boundary TEE.

Let us now analyze the geodesic structure (\ref{CWES3}). In this case, the space-like geodesics $A_1A_2$ and $B_1B_2$ are respectively described by Eqs.~(\ref{spacelike1CWEScase1}) and (\ref{spacelike2CWEScase1}), whereas the time-like geodesic $A_2B_2$ is described by Eq.~(\ref{timelikeCWEScase2}). Taking the bulk coordinates of points $A_2$ and $B_2$ as
\begin{eqnarray}
& & A_2=(t_2,z_2),~~~B_2=(t_3,z_3),~~~\text{with}~~t_2>t_3\,,
\end{eqnarray} 
the length of the time-like geodesic is again given by (\ref{lengthtimelikeCWES2}), with joints appearing at null infinities, i.e., $t_2=-t_3=\infty$. Accordingly, the length of the time-like geodesic $A_2B_2$ is 
\begin{eqnarray}
 \mathcal{A}_t & = & i L \int_{-\infty}^{\infty} dt \frac{R^\alpha}{\alpha^2 (t-t_0)^2 + R^{2\alpha}} = \frac{i L}{\alpha} \tan^{-1} \left(\frac{\alpha (t-t_0)}{R^\alpha} \right)\bigg\rvert_{-\infty}^{\infty},\\ \nonumber
 & = & \frac{i \pi L}{\alpha}\,.
\label{lengthtimelikeCWES3}
\end{eqnarray}
To find $t_0$, like before, we expand Eqs.~(\ref{spacelike1CWEScase1}) and (\ref{timelikeCWEScase2}) near the future null infinity. This gives us
\begin{eqnarray}
& & z^\alpha = \alpha(t-t_0)+\mathcal{O}(1/t),~~~t=\frac{z^\alpha}{\alpha}-\frac{C_{+}}{\alpha}+\frac{T}{2}+\mathcal{O}(1/z^\alpha)\,.
\end{eqnarray}
These two equations represent the same point $A_2$ when $z\rightarrow \infty$. Therefore, we have $t_0=-C_{+}/\alpha+T/2$. Doing the analogous expansion for the point $B_2$, we get $t_0=C_{-}/\alpha - T/2$. Accordingly, we have the relation
\begin{eqnarray}
& & C_{+} + C_{-}=\alpha T\,.
\end{eqnarray}
With this relation of $C_{\pm}$ in hand, we can now compute the total length of the space-like geodesics $A_1A_2$ and $B_1B_2$
\begin{eqnarray}
\mathcal{A}_s & = &  L \left[ \int_{\epsilon}^{z_2=\infty} dz~ \frac{C_{+}}{z\sqrt{z^{2\alpha}+C_{+}^{2}}} + \int_{\epsilon}^{z_3=\infty} dz~ \frac{C_{-}}{z\sqrt{z^{2\alpha}+C_{-}^{2}}}  \right]\,,  \nonumber \\
 & = & \frac{L}{\alpha} \log{\left( \frac{4 C_{+} C_{-}}{\epsilon^{2\alpha}}   \right)} = \frac{L}{\alpha} \log{\left( \frac{4 C_{+} (\alpha T - C_{+})}{\epsilon^{2\alpha}}\right)}
\end{eqnarray}
The joint condition $\partial A_s/\partial C_{+}=0$ further gives us the relation $C_{+}=\alpha T/2$. Therefore,
\begin{eqnarray}
\mathcal{A}_s & = &  \frac{2 L}{\alpha} \log{\left( \frac{\alpha T}{\epsilon^{\alpha}}\right)}\,.
\label{CWEStotalspacelengthcase3}
\end{eqnarray}
Since the geodesic stricture (\ref{CWES3}) is a consistent CWES, the corresponding TEE from the proposal (\ref{TEEproposalCWES}) is  
\begin{eqnarray}
 S_{A}^{T} & = & \frac{1}{4 G_3}\left( \mathcal{A}_s + \mathcal{A}_t  \right)\,,\\
 & = &  \frac{L}{4 G_3}\left( \frac{2 }{\alpha}\log{\frac{\alpha T}{\epsilon^\alpha}} +  i \frac{\pi }{ \alpha}  \right)\,.
\label{TEEM3}
\end{eqnarray}
which is exactly the TEE expression obtained in the last section. In fact, we can also trace back the reason for this equivalence. In particular, note that for $C_{\pm}=\alpha T/2$, which is obtained from the joint equation of CWES, the space-like geodesic equation of CWES reduces to
\begin{eqnarray}
t=\pm\frac{1}{\alpha}\sqrt{z^{2\alpha}+\frac{\alpha^2 T^2}{4}}\,.
\end{eqnarray}
This is precisely the geodesic equation used in the last section, thereby giving equal results for the geodesic lengths and the TEE. 

It is quite interesting to see that the CWES formalism and joint conditions pick out a particular value of constants and, hence, a minimal bulk surface which is homologous to the boundary subsystem. This was shown to be true for pure AdS$_3$ background in \cite{Li:2022tsv}, and here we show that it is also true for three-dimensional anisotropic Lifshitz background. Since the CWES formalism reduces to the usual Ryu-Takayanagi entanglement entropy formalism for space-like subsystems and the fact that it also allows us to compute the minimal surfaces unambiguously for time-like intervals indeed puts weight behind the correctness of this formalism. However, it seems that this CWES formalism is more efficient when analytic results for the bulk surfaces are available. In many cases, in particular, in higher-dimensional holographic theories, analytic results for the bulk entangling surfaces are not known. In those cases, implementing the joint conditions and finding CWES would not be as straightforward as in three dimensions, thereby restricting the potential applicability of the CWES formalism. In any case, it would be interesting to see if this formalism can be applied to compute TEE at finite temperatures in four and higher dimensions or in non-conformal field theories. We leave this interesting exercise for future work.

\section{HTEE via Smooth Merging: Method~3}
\label{sec4}
In this section, we compute the holographic TEE in the Lifshitz field theory using the formalism of \cite{Afrasiar:2024lsi}. In this formalism, the main idea is to smoothly merge the time-like and space-like minimal surfaces so that their union is homologous to the boundary and has a well-defined first derivative at the merging point. This formalism further imposes a connection between the integration constants of the time-like and space-like minimal surfaces. In particular, the integration constant of the space-like minimal surface is assumed to be fixed once the corresponding integration constant of the time-like minimal surface is fixed. Here, we closely follow the computation of \cite{Afrasiar:2024lsi}, however, now in the Lifshitz background.

To evaluate TEE using the formalism of \cite{Afrasiar:2024lsi}, we start with the minimal area integral
\begin{eqnarray}
& & S_{A}^{T} = \frac{L}{4 G_3} \int~dz \sqrt{\frac{1}{z^2}-\frac{t'(z)^2}{z^{2\alpha}}}.
\label{TEEM3}
\end{eqnarray}
The above action leads to the following equation of motion for the minimal surfaces,
\begin{eqnarray}
& & t'(z)= \pm \frac{C z^{2\alpha-1}}{\sqrt{1+ C^2 z^{2 \alpha}}}\,,
\label{tEOMM3}
\end{eqnarray}
where $C$ is an integration constant. Notice that depending upon whether $C^2$ is positive or negative there can be two different types of extremal surfaces. In particular, for $C^2<0$, the denominator of Eq.~(\ref{tEOMM3}) can be zero, leading to a surface exhibiting a turning point at $z=z_*$. This allows us to fix the integration constant in terms of the turning point $C^2=-z_{*}^{-2\alpha}$. At the turning point, we also have $t'(z_*)=\infty$. For $C^2>0$, it is obvious that the surface corresponding to Eq.~(\ref{tEOMM3}) does not exhibit a turning point. We will see shortly that the $C^2<0$ surface appears only near the IR boundary and gives an imaginary value whereas the $C^2>0$ surface goes from the asymptotic boundary to the deep IR and yields real value. Following \cite{Afrasiar:2024lsi}, we call these surfaces $t_{Im}(z)$ and $t_{Re}(z)$ respectively. Note that once the constant $C$ is fixed for the time-like surface, it is amused to be fixed for the space-like surface as well, i.e., 
\begin{eqnarray}
C^2 & = & -z_{*}^{-2\alpha} < 0,~~~\text{for time-like}\,\\
C^2 & = & z_{*}^{-2\alpha} > 0,~~~\text{for space-like}\,.
\end{eqnarray}

Next, the hypersurfaces comprising the TEE are defined
\begin{eqnarray}
& & \Sigma_{Im}:~~~ t-\int dz~t_{Im}'(z)=0\,,\\
& & \Sigma_{Re}:~~~ t-\int dz~t_{Re}'(z)=0\,.
\end{eqnarray}
It is straightforward to see that $\Sigma_{Im}$ and $\Sigma_{Re}$ are indeed time-like and space-like surfaces in their respective domain. The total boundary length is determined by merging the space-like and time-like surfaces smoothly in such a way that the resultant surface is homologous to the boundary and has a well-defined first derivative at the merging point. For the Poincar\'{e} patch, the merging occurs at $z=\infty$. At the merging point, we have $t_{Re}'(\infty)=t_{Im}'(\infty)=\pm z^{\alpha-1}$ and $g^{\mu\nu}\partial_{\mu}\Sigma_{Im}\partial_{\nu}\Sigma_{Im}=z_{*}^{2\alpha}=-g^{\mu\nu}\partial_{\mu}\Sigma_{Re}\partial_{\nu}\Sigma_{Re}$. 

\begin{figure}[htb!]
    \centering
    \begin{tikzpicture}[scale=0.6]

\draw[thick, ->] (1,0) -- (12,0) node[right] {$z$};
\draw[thick, ->] (11,-7) -- (11,7) node[left] {$t$};

\coordinate (A) at (1,4);
\coordinate (B) at (1,-4);
\coordinate (zs) at (4,0);
\coordinate (T1) at (9,6);
\coordinate (T2) at (9,-6);
\node[right] at (4,0.25) {$z_*$};
\node[right] at (T1) {$\frac{T}{2}$};
\node[right] at (T2) {$-\frac{T}{2}$};
\node[below] at (1,7.5) {\textcolor{teal}{$z=\infty$} };

\draw[blue, thick] (T1) to[out=180, in=90, looseness=0.5] (A);
\draw[blue, thick] (T2) to[out=180, in=-90, looseness=0.5] (B);
\draw[red, thick] (A) to[out=-90, in=90, looseness=0.5] (zs) to[out=-90, in=90, looseness=0.5] (B);
\draw[teal,dashed] (1,-7) -- (1,7) ;
\end{tikzpicture}
    \caption{Pictorial representation of merging of space-like and time-like surfaces at $z=\infty$. Here blue lines correspond to the space-like surface and red lines correspond to the time-like surface.}
    \label{geodesicsmethod3}
\end{figure}
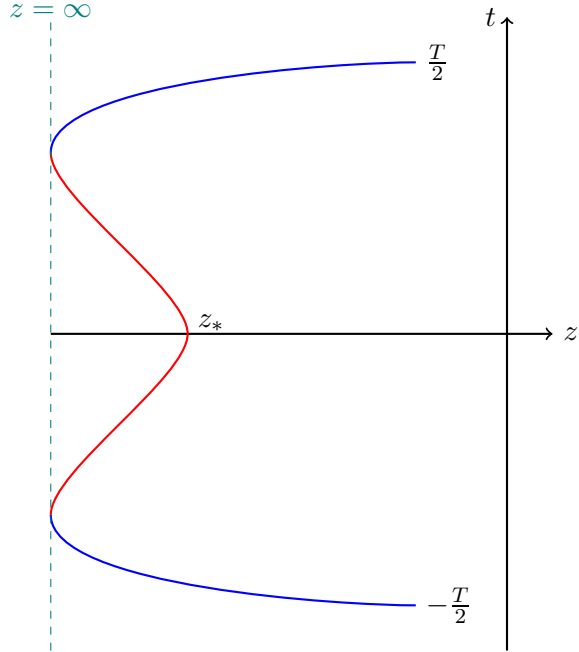

Essentially, putting all together, in the formalism of \cite{Afrasiar:2024lsi} the TEE is given by the combination of two space-like surfaces that go from the asymptotic boundary at $z=0$ to $z=\infty$ and two time-like surfaces that go from $z=\infty$ to a turning point $z=z_*$. These surfaces are then merged at $z=\infty$. A pictorial representation of these surfaces and their merging is shown in Fig.~\ref{geodesicsmethod3}.  

Since the merging is done at $z=\infty$ at which $t_{Im}'(\infty)=t_{Re}'(\infty)$ and $g^{\mu\nu}\partial_{\mu}\Sigma_{Im}\partial_{\nu}\Sigma_{Im}=z_{*}^{2\alpha}=-g^{\mu\nu}\partial_{\mu}\Sigma_{Re}\partial_{\nu}\Sigma_{Re}$, the boundary length of the surface is then given by
\begin{eqnarray}
& & T = 2\int_{z_*}^{\infty} dz~t_{Im}'(z) - 2\int_{0}^{\infty} dz~t_{Re}'(z) = \frac{2z_{*}^\alpha}{\alpha}\,.
\label{lengthM3}
\end{eqnarray}
We now evaluate the areas of these surfaces to compute TEE. For the space-like surface with $C^2=z_{*}^{-2\alpha}$, we have 
\begin{eqnarray}
 S_{A}^{T_{Re}} & = & \frac{2 L}{4 G_3}\int_{\epsilon}^{\infty}~dz \sqrt{\frac{1}{z^2}-\frac{t_{Re}'(z)^2}{z^{2\alpha}}}= \frac{2 L}{4 G_3}\int_{\epsilon}^{\infty}~dz \frac{1}{z\sqrt{1+C^2 z^{2\alpha}}}   \\
& = & \frac{L}{2 G_3 \alpha}\log{\frac{2 z_{*}^\alpha}{\epsilon^\alpha}}\,.
\label{TEERe1M3}
\end{eqnarray}
Substituting Eq.~(\ref{lengthM3}) into Eq.~(\ref{TEERe1M3}), we get
\begin{eqnarray}
& & S_{A}^{T_{Re}} = \frac{L}{2 G_3 \alpha}\log{\frac{\alpha T}{\epsilon^\alpha}}\,.
\label{TEERe2M3}
\end{eqnarray}
Similarly, for the time-like surface with $C^2=-z_{*}^{-2\alpha}$, we have
\begin{eqnarray}
& & S_{A}^{T_{Im}} = \frac{2 L}{4 G_3}\int_{z_{*}}^{\infty}~dz \sqrt{\frac{1}{z^2}-\frac{t_{Im}'(z)^2}{z^{2\alpha}}}=\frac{L}{4 G_3 }\frac{i \pi }{\alpha}\,.
\label{TEEIm1M3}
\end{eqnarray}
Combining (\ref{TEERe2M3}) and (\ref{TEEIm1M3}), we get the TEE
\begin{eqnarray}
& & S_{A}^{T} = S_{A}^{T_{Re}} + S_{A}^{T_{Im}} =\frac{L}{4 G_3}\left( \frac{2}{\alpha}\log{\frac{\alpha T}{\epsilon^\alpha}} +  i \frac{\pi }{ \alpha}  \right)\,.
\label{TEE1M3}
\end{eqnarray}
The above expression of $S_{A}^{T}$ agrees with the analogous expression obtained for TEE from other methods in previous sections.

We end this section by discussing a few technical issues and observations in the above computation of TEE using the formalism of \cite{Afrasiar:2024lsi}. Note that this formalism requires the constant $C$ [see Eq.~(\ref{tEOMM3})] to be fixed for the space-like surface once it is fixed for the time-like surface, i.e., $C^2=\pm z_{*}^{2 \alpha}$ ($+$ for space-like and $-$ for time-like). However, there is no physical argument that we have to follow such a choice, and the constant, in fact, can be different for different surfaces. Indeed, note that different constants appeared for space-like and time-like geodesics in the last sections. Therefore, imposing such a constraint on the constant seems a bit forced in the formalism of \cite{Afrasiar:2024lsi}. This could be a cause of concern considering that the turning point does not appear in the area integral of the space-like surface unless we take $C^2=z_{*}^{2\alpha}$. Indeed, one expects that the area of the space-like surface should not depend on the turning point if there is no turning point in it. We feel that it is an important issue that needs to be discussed further.

It is interesting to ask if any relation exists between the constant $C$ here and the constants that appeared in the geodesic equations in the previous sections. It is clear that a relation might exist for the time-like case. A straightforward comparison between (\ref{timelikelengthM1}) and (\ref{TEEIm1M3}) shows $R=z_*$, suggesting that the arbitrary constant $R$ in the time-like geodesic is related to its turning point. 

Though the derivatives of real and imaginary surfaces are equal, they diverge at the merging point, i.e., $t_{Re}'(\infty)=t_{Im}'(\infty)=\pm z^{\alpha-1}$. This was not the case for AdS$_3$. Indeed, notice that for $\alpha=1$, the derivatives just become one. However, they diverge for any finite $\alpha$, suggesting that derivatives are not well defined at the merging point. Moreover, it is easy to see that the diverging nature of the derivatives of $t_{Re}$ and $t_{Im}$ at the merging point will persist for any spacetimes that have anisotropic scaling between space and time coordinates, such as the higher-dimensional Lifshitz theories. However, this issue may not arise for pure AdS type geometries.

\section{HTEE via coordinate complexification: Method 4}
\label{sec5}
In this section, we compute the holographic TEE in the Lifshitz field theory using the formalism suggested in \cite{Heller:2024whi}. This formalism introduces a novel idea to complexify the spacetime coordinates to produce real and imaginary parts of TEE and suggests that the bulk carrier of the holographic TEE consists of codimension-two extremal surfaces $\Gamma_T$, which are anchored to a time-like boundary subregion $A$ and typically extend into a complexified bulk geometry. In this formalism, the holographic TEE is given by
\begin{equation}
    S_{A}^{T} = \frac{\mathcal{A}(\Gamma_T)}{4G_3}\,,
    \label{area}
\end{equation}
where $\mathcal{A}$ is the area of $\Gamma_T$ in the complexified bulk geometry. For our case, we take the following parametrization of the codimension-two bulk extremal surface
\begin{equation}
    X^\mu(\lambda)=\{t(\lambda),z(\lambda),x=0\}\,,
    \label{extremalsurface}
\end{equation}
where $\lambda$ represents a parameter that varies along the surface. With the above parametrization, we need to extremize the following area functional
\begin{equation}
    \mathcal{A}= \int d\lambda \mathcal{L} =L\int d\lambda\sqrt{ \frac{-t'(\lambda)^2 z(\lambda)^2+ z'(\lambda)^2 z(\lambda)^{2 \alpha}}{z(\lambda)^{2 \alpha+2}}}\,,
    \label{areadensity}
\end{equation}
to compute TEE. Because the bulk metric Eq.~(\ref{lifshitzmetric}) is time-independent, the surface $\Gamma_T$ has a conserved quantity, $P$. This simplifies the Euler-Lagrange equation derived from Eq.~(\ref{areadensity}) to a first-order form
\begin{equation}
    t'(\lambda)^2 =\frac{P^2  z(\lambda)^{-2 + 4 \alpha} z'(\lambda)^2}{1 + P^2  z(\lambda)^{2 \alpha}}\,.
    \label{tprime}
\end{equation}
Eq.~(\ref{tprime}) shows that the point $z=z_*$, where $1 + P^2  z_{*}^{2 \alpha}=0$, corresponds to the turning point of $\Gamma_T$, where a branch-point singularity occurs in $t(z)$. Interestingly, one may think of the turning point condition $P^2=-z_{*}^{-2\alpha}$ as a complexified version of turning point condition $C^2=-z_{*}^{-2\alpha}$ that appeared for the time-like geodesic in the last section.

In three dimensions, $\Gamma_T$ is a bulk geodesic anchored at the boundary. We use $\lambda$ as an affine parameter with $\mathcal{L}=1$, ensuring that the bulk geodesic reaches the asymptotic boundary at $\lambda=\pm \lambda_*$,
\begin{equation}
    z(\pm \lambda_*) = \epsilon, \hspace{10pt} t(\pm \lambda_*) = \pm \frac{T}{2}\,,
    \label{boundaryconditions}
\end{equation}
where $\epsilon$ is again a UV cut-off. The solutions to Eq.~(\ref{tprime}), under the boundary conditions outlined in Eq.~(\ref{boundaryconditions}) is expressed as,
\begin{equation}
    t(\lambda) = \sqrt{ \frac{\alpha^2 T^2}{4} - \epsilon^{2 \alpha}} \frac{\tanh(\alpha\lambda)}{\alpha}, \hspace{10pt} z(\lambda)=\left(i \frac{\sqrt{ \frac{\alpha^2 T^2}{4} - \epsilon^{2 \alpha}}}{\cosh(\alpha \lambda)} \right)^{\frac{1}{\alpha}}\,.
\end{equation}
It is interesting to see that even for the Lifshitz geometry, analytic expressions of $t(\lambda)$ and $z(\lambda)$ can be found. For small $\epsilon$, the $\lambda_*$ value will be,
\begin{equation}
    \lambda_*= \frac{1}{\alpha}\log \frac{\alpha T}{ \epsilon^{ \alpha} } + \frac{i \pi}{2 \alpha}\,.
\end{equation}
The timelike entanglement entropy for a region in the Lifshitz geometry is captured by the regularized geodesic length, which equals $2 \lambda_*$. This gives
\begin{eqnarray}
   S_{A}^{T} =  \frac{L}{4 G_3} \left(\frac{2}{\alpha}\log \frac{\alpha T}{ \epsilon^{ \alpha} } + \frac{i \pi}{ \alpha}\right)\,,
\end{eqnarray}
which again matches with TEE expressions obtained by other methods in previous sections.

\section{Conclusion and discussion}
\label{sec6}
In this paper, we have explored the holographic TEE in the boundary theory of three-dimensional Lifshitz spacetimes. By employing four distinct techniques -- geodesics, complex-valued extremal surfaces (CWES), smooth merging of time-like and space-like geodesics, and coordinate complexification -- we investigated the structure of TEE in Lifshitz spacetimes, which are characterized by anisotropic scaling between time and space coordinates. In the first three techniques, the TEE consisted of joint space-like and time-like surfaces, whereas the last technique is based on the complexification of spacetime coordinates.

We obtained the analytic result for TEE in each formalism, compared the technique, and analyzed how the anisotropic scaling affects the TEE. We found that all these holographic formalisms give the same results for TEE in the Lifshitz background as well. We found that both the real and imaginary parts of TEE depend on $\alpha$, indicating that different Lifshitz theories characterized by a different anisotropic parameter $\alpha$ exhibit different TEE structures. Similar to the CFT$_2$ case, the imaginary part of TEE does not depend on subsystem length but relies on $\alpha$. The TEE expression of the Lifshitz theory reduces to the CFT$_2$ expression in the limit $\alpha\rightarrow 1$. In the case of AdS$_3$, the correctness of the holographic TEE result can be verified from the independent analytic continuation technique. However, unfortunately, no independent result is available for TEE in two-dimensional anisotropic Lifshitz field theory to compare our holographic results.

The holographic space-like entanglement entropy for a Lifshitz theory is given by:
\begin{equation}
    S_A = \frac{L}{2 G_3} \log \left(\frac{l}{\epsilon}\right)\,,
    \label{LifshitzEE}
\end{equation}
where $l$ represents the length of the space-like subsystem. This expression is obtained from the Ryu-Takayanagi prescription. An analytic continuation using a Wick rotation has been proposed to extend this concept to a time-like entanglement entropy \cite{Doi:2023zaf}. In this approach, the time-like entanglement entropy $S_{A}^{T}$ can be derived by substituting the subsystem interval $l$ with the temporal interval $T$. This substitution is made as $l \rightarrow (i \alpha T)^{1/\alpha}$, effectively applying a transformation that aligns with the results of our time-like entanglement entropy $S_{A}^{T}$.

However, there are a few subtle issues that need to be emphasized in the context of space-like EE in the Lifshitz theory and one needs to be careful in obtaining TEE from analytic continuation from space-like EE.  Note that the above expression of the space-like EE in the Lifshitz theory matches the conformal field theory expression in two dimensions. This is because the Ryu-Takayanagi prescription for a fixed time slice does not depend on the $g_{tt}$ component. Accordingly, two metrics that differ only in their $g_{tt}$ components are bound to give the same expression for the space-like EE holographically. However, this equality breaks down when the system has higher derivative terms and one needs to extremize the EE action containing Ricci tensors and extrinsic curvatures \cite{Hung:2011xb, Dong:2013qoa}. Interestingly, one can also obtain Lifshitz spacetimes as solutions to massive gravity theory containing higher derivative terms \cite{Bergshoeff:2009hq, Deser:1982vy}. In \cite{Hosseini:2015gua}, it was found that higher derivative terms inspired Lifshitz spacetime gives $\alpha$ dependent space-like EE which is different from Eq.~(\ref{LifshitzEE}). The analytic continuation of the form $l \rightarrow (i \alpha T)^{1/\alpha}$ is then expected to give a different TEE expression compared to the one obtained in previous sections. 

Similarly, in holographic models inspired by cMERA, the space-like entanglement entropy for a massless Lifshitz field theory with a general dynamical exponent $\alpha$ is given by \cite{Nozaki:2012zj,He:2017wla,Gentle:2017ywk}
\begin{equation*}
    S_A = \frac{\alpha}{3} \log \left(\frac{l}{\alpha \epsilon}\right)\,.
\end{equation*}
This expression again contrasts with the EE obtained directly from the Lifshitz vacuum, suggesting subtle differences between standard holographic methods and cMERA-based approaches. These differences may arise due to the unique way cMERA captures the non-relativistic scaling symmetries of Lifshitz theories. 

Moreover, in holography, the two-point functions of the boundary operators are given by geodesics in the bulk. Since, in three dimensions the extremal surfaces in the Ryu-Takayanagi prescription correspond
to the geodesic in the bulk, it allows us to relate EE with the two-point function. The same for Lifshitz field theories can be determined by symmetry constraints, similar to how it's done in CFTs through Ward identities, though with some key differences due to the Lifshitz scaling symmetry, see for instance \cite{Keranen:2016ija, Keranen:2012mx, Park:2022mxj}. In standard CFTs, the full conformal symmetry fixes the form of the two-point function. In a Lifshitz theory, the two two-point functions explicitly depend on the anisotropic exponent $\alpha$. For an operator $\mathcal{O}(t,x)$ with scaling dimension $\Delta$, the lack of Lorentz symmetry in Lifshitz field theories leads to two-point functions taking distinct forms in the spatial and temporal directions. The spatial two-point function is given by, 
\begin{equation*}
    \langle \mathcal{O}(t,x_1) \mathcal{O}(t,x_2) \rangle = \frac{1}{|x_1-x_2| ^{2 \Delta} }\,,
\end{equation*}
and the temporal two-point function becomes,
\begin{equation*}
    \langle \mathcal{O}(t_1,x) \mathcal{O}(t_2,x) \rangle = \frac{1}{|t_1-t_2| ^{2 \Delta/\alpha} } \,.
\end{equation*}
One can similarly obtain the two-point function for general space-time separated points. This is needed for analytic continuation from the space-like interval to the time-like interval. However, even with the two-point function in hand, we might not be able to say anything conclusively for TEE via analytic continuation from space-like EE. In particular, depending on the Lifshitz system and its vacuum, different space-like EE expressions have been suggested in the literature, see for instance \cite{Basak:2023otu, Gentle:2017ywk, Hosseini:2015gua}. The analytic continuation is then bound to give different expressions for TEE. Therefore, the use of analytic continuation to get TEE is not as straightforward as it seems for the Lifshitz system and more work is indeed needed in this direction. Further investigation into these nuances could deepen our understanding of entanglement structures in Lifshitz field theories and refine holographic models for such systems. We hope to say something on this issue in the near future.

Overall, this work provides a small contribution to the growing body of research on TEE. Our analysis can guide future studies into the role of TEE in Lifshitz-type theories. For future work, a natural extension of this study would be to explore TEE in higher-dimensional Lifshitz backgrounds and see if the results of different holographic techniques again correlate well. The TEE was computed in a confining geometry using the formalism of \cite{Afrasiar:2024lsi} and it was observed that TEE can act as a probe for confinement like its spatial counterpart \cite{Klebanov:2007ws, Dudal:2016joz, Dudal:2018ztm, Mahapatra:2019uql, Jain:2020rbb, Jain:2022hxl, Arefeva:2020uec}. It would be interesting to explore TEE in various other top-down and bottom-up confining backgrounds using the formalism of \cite{Afrasiar:2024lsi} as well as the formalism of \cite{Heller:2024whi}. We leave these interesting exercises for future work.

\section*{Acknowledgements}
We would like to thank I.~Banerjee for the useful discussion. The work of S.~S.~J.~is supported by Grant No. 09/983(0045)/2019-EMR-I from CSIR-HRDG, India. The work of S.~M.~is supported by the core research grant from the Science and Engineering Research Board, a statutory body under the Department of Science and Technology, Government of India, under grant agreement number CRG/2023/007670. 

 \bibliographystyle{JHEP}
\bibliography{mybib}
\end{document}